\documentclass[runningheads]{llncs}

\usepackage[colorlinks,urlcolor=blue,linkcolor=blue,citecolor=blue]{hyperref}

\usepackage{graphicx}

\usepackage{comment}

\usepackage{array}

\author{Neil P. Chue Hong$^1$, Jeremy Cohen$^2$, Caroline Jay$^3$}
\date{%
    $^1$University of Edinburgh\\
    $^2$Imperial College London\\
    $^3$University of Manchester\\
    ~\\
    \today
}

\begin{document}


\title{Understanding Equity, Diversity and Inclusion Challenges Within the Research Software Community}

\titlerunning{Understanding EDI Challenges Within the Research Software Community}
%

\author{Neil P. Chue Hong\inst{1}\orcidID{0000-0002-8876-7606} \and
Jeremy Cohen\inst{2}\orcidID{0000-0003-4312-2537} \and
Caroline Jay\inst{3}\orcidID{0000-0002-6080-1382}}
\authorrunning{N.P. Chue Hong et al.}
%
\institute{Software Sustainability Institute \& EPCC, University of Edinburgh, Edinburgh, U.K.\\
\email{N.ChueHong@epcc.ed.ac.uk}\\
\and
Department of Computing, Imperial College London, London, U.K.\\
\email{jeremy.cohen@imperial.ac.uk}\\
\and
Department of Computer Science, University of Manchester, Manchester, U.K.\\
\email{Caroline.Jay@manchester.ac.uk}\\
}

\maketitle

\begin{abstract}

Research software -- specialist software used to support or undertake research -- is of huge importance to researchers. It contributes to significant advances in the wider world and requires collaboration between people with diverse skills and backgrounds. Analysis of recent survey data provides evidence for a lack of diversity in the Research Software Engineer community. We identify interventions which could address challenges in the wider research software community and highlight areas where the community is becoming more diverse. There are also lessons that are applicable, more generally, to the field of software development around recruitment from other disciplines and the importance of welcoming communities. 

\keywords{research software, software engineering, research software engineering, diversity, EDI.}

\end{abstract}

\section{Introduction}

Developing specialist research software to support computational science is an especially challenging process. Unlike more traditional software engineering tasks, researchers or, increasingly, RSEs (Research Software Engineers) who write this software need an understanding of the underlying scientific challenge being addressed. 

The methods used in research favour a hypothesis-driven approach. This creates a different working environment from the wider software engineering industry, where software is built to meet a client's specification. As noted by Hettrick et al., 
\begin{quote}
``\textit{To be effective, software development in research should be approached, not as a one-off transaction, but as a partnership between researcher and software expert.}'' \cite{RSESOTN2017}
\end{quote}

The ``partnership between researcher and software expert'' mentioned here highlights a need for varied skills and good communication. It is clear that much computational research requires diversity of skills and experience, working in partnership, that spans software engineering and science. Researchers are more likely to be working with others who have different technical expertise, use different technical terminology, and may be communicating in a tertiary language.

Diversity in project teams, and workplaces in general, is important and it is widely accepted that there are many benefits from ensuring diversity within teams and communities. For example, there is evidence to show that diversity in terms of knowledge or skills can be beneficial~\cite{Liang_2007} and that gender diversity within groups can result in higher quality scientific outputs~\cite{Campbell2013}. However, there are cases where diversity can raise challenges, for example with reduced feelings of well-being amongst members in highly-diverse teams~\cite{Van_Der_Zee_2004}. Diversity in the context of teams in industry and research is an area that has been the subject of extensive research and there are some contradictory results from the many studies undertaken. Stahl et al.~\cite{Stahl_2009} look at a number of different studies of cultural diversity concluding that there are both benefits and drawbacks. Whether benefits can be realised while minimising more challenging aspects will depend on effective process management. In the context of software engineering, Capretz \& Ahmed~\cite{Capretz_2010} looked at how different personality traits relate to suitability for different roles in software projects. Ultimately they conclude that diversity in terms of personalities and skills is important in helping with problem solving tasks involved in building and maintaining software.

The different types of diversity highlighted so far, including diversity of skills and knowledge, culture, personality and gender are, of course, just some of the many different aspects that lead to diversity amongst a group of individuals. Others, among a huge range, include ethnicity, disability and age. 

Research Software Engineers are, at present, much more likely to come from a research background than software engineering professionals working in other fields. Nonetheless, we know from surveys within the RSE community that a similar diversity crisis to that identified in other fields exists. In this paper, we examine current problems with diversity in research software engineering and consider potential causes. Our aim is to present a better understanding of existing diversity within the RSE community in order to develop insights and recommendations to address current issues. We achieve this through an empirical analysis of existing open data collected and made available through large-scale international surveys of research software engineers undertaken by the RSE community. We also review related work from allied areas which suggests ways of addressing the challenges identified.

As pointed out by Mathieson in~\cite{DiversityCreativitySE}, based on 2018 ONS data, just one in eight of almost 340,000 software development professionals in the UK are women. This example demonstrates that there is a long way to go to address the lack of diversity in the wider software engineering domain, as well as the more focused domain of research software engineers. Professional software engineers can gain their skills through a variety of different routes and they may come from a wide range of different disciplines. For example, they may have developed their skills through a degree programme or vocational training in computing or a related area. Alternatively, they may come from a completely different disciplinary background and have re-trained through one of a large number of ``coding schools'' that offer training, often via intensive courses, in software development skills. The wide array of routes into the field creates an expectation of a diverse field which makes it all the more surprising that this is not the case.

We highlight three core contributions this paper provides to understanding the importance of diversity among RSEs working in computational and data science:
\begin{itemize}
\item Providing an evidence-based analysis that demonstrates the current problems with diversity in the research software engineering community.
\item Demonstrating that there is already extensive ``domain mobility'' for RSEs and that a lack of such mobility is therefore not likely to be a cause for a lack of diversity within the field.
\item Offering four general recommendations that we believe can form a basis to support addressing the lack of diversity in research software engineering.
\end{itemize}

In Section~\ref{section:survey} we examine the International RSE Survey results from 2018 and undertake further analysis on this data. Section~\ref{section:improving-diversity} highlights the three areas that we see as both helping to explain and provide the basis for addressing the lack of diversity amongst RSEs, while discussion and conclusions are provided in Section~\ref{section:conclusion}.

\section{Surveying RSE Diversity}
\label{section:survey}

To understand the diversity challenges facing those embarking on careers as Research Software Engineers, we require a better understanding of the landscape as a whole. The Software Sustainability Institute coordinates an ongoing series of surveys of the RSE community. In the most recent survey \cite{RSESurvey2018}, from 2018, participants were asked a number of socio-demographic questions. 

\begin{table}
\vspace{-4mm}
\centering
\begin{tabular}{ || m{3cm} | m{5em} | m{7em} | m{7em} | m{7em} || } 
 \hline \hline
Percentage of UK who are: & RSEs\cite{RSESurvey2018} & Academics\cite{HESA:Table-3} & Software Developers\cite{BCSDiversity2020} & All UK workers\cite{BCSDiversity2020} \\
 \hline  \hline
 Gender (female) & 14 & 46 & 14 & 48 \\
  \hline
 Ethnicity (BAME/Mixed) & 5 & 15 & 21 & 12 \\
 \hline
 Report disability & 6 & 4 & 10 & 13 \\
 \hline \hline
\end{tabular}
\caption{Comparison of 2018 demographics  for Research Software Engineers, Academics, Software Developers and general working population in the United Kingdom.}
\label{table:table0}
\vspace{-8mm}
\end{table}

Our reanalysis primarily focuses on UK data, as this is where the authors are located, but the numbers from other countries in the survey are broadly similar. 
Table \ref{table:table0} compares RSEs with UK Higher Education Statistics Authority (HESA) data \cite{HESA:Table-3}, and a study by the British Computer Society (BCS) of  software development professionals in the UK \cite{BCSDiversity2020} based on 2018 Office for National Statistics data. 
Limitations in the RSE data currently available to us mean that we focus on aspects of gender, ethnicity and disability, as opposed to other types of diversity.

This comparison indicates a gender diversity gap, which we might expect given the percentage of women working as software developers is also 14\%. But this could be better: software is a fundamental part of all research, and 46\% of academic staff are female. 
The ethnicity data indicates a greater problem. Of the respondents who declared their ethnicity, there were only 6 non-white and 5 mixed race RSEs -- 5\%. 
These figures are significantly lower than the 28\% BAME students studying computer science \cite{UUKtrends2018} and 10\% of BAME physicists \cite{IOPPhysicist}. Although we should not aggregate non-white ethnic categories together, as they will experience different challenges and biases, it is striking that RSEs do not fit the general profile of those working in the IT industry, where there is greater ethnic diversity.

\begin{figure}[h]
\vspace{-8mm}
    \center{\includegraphics[width=0.75\linewidth]{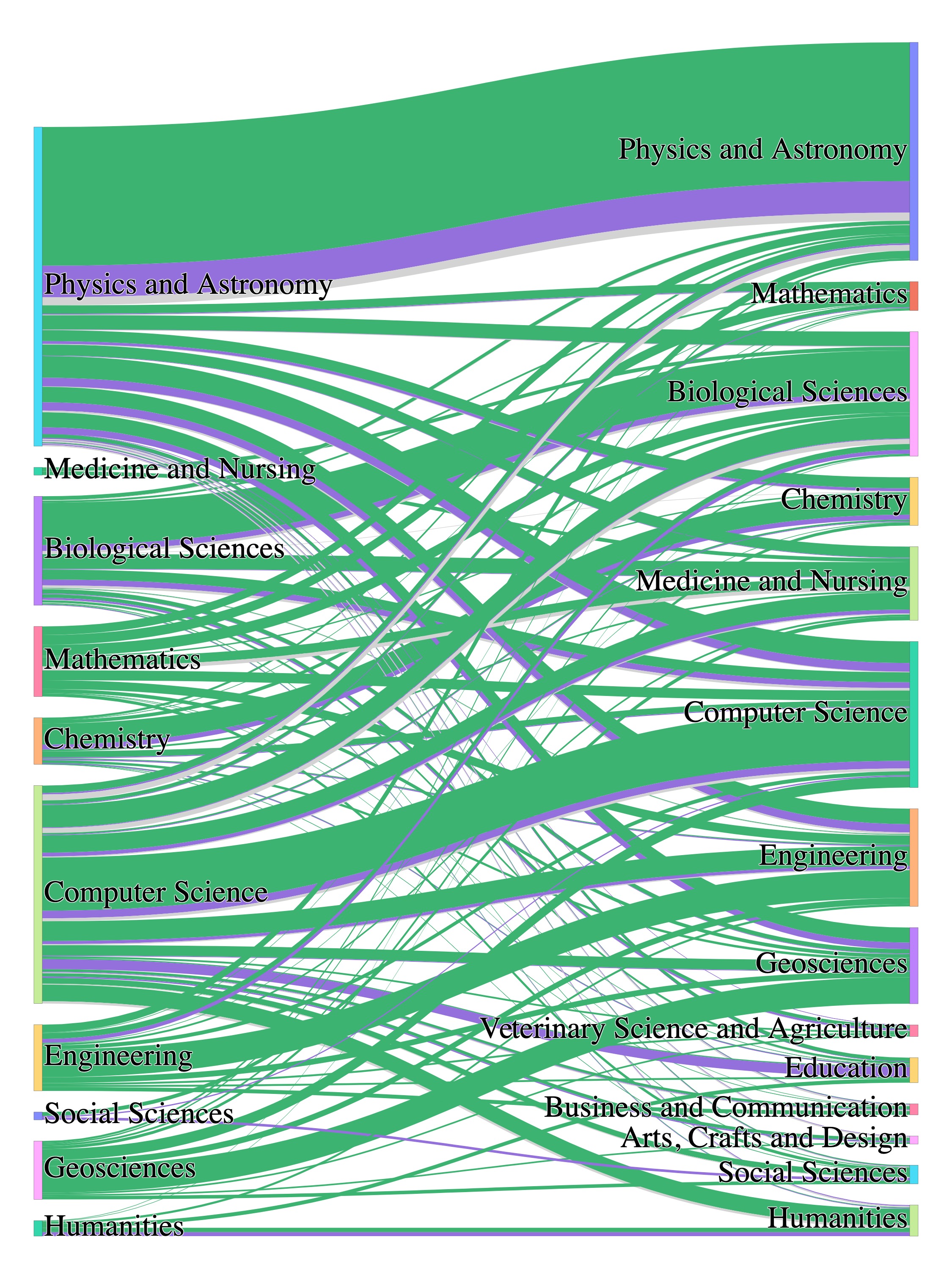}}
    \caption{\label{fig:rse-flow} Flow of UK RSEs from academic discipline studied to domains they now work in, categorised by gender: 79.8\% male (green) / 14.3\% female (purple) / 5.9\% other responses (grey). N=203.}
\vspace{-8mm}
\end{figure}

An important area of research is understanding if diversity can be improved if RSEs are drawn from a wider set of degree backgrounds. Figure~\ref{fig:rse-flow} shows where RSEs are working based on their academic degree. Only gender diversity was considered, given the low ethnic diversity and number of respondents reporting a disability. Gender balance is relatively uniform across fields rather than mirroring the gender balance in that field. Over half of RSEs have a first degree in Physics and Astronomy or Computer Science. In the UK, 17\% of CS undergraduates were female compared with 41\% of physical sciences undergraduates \cite{HESA:Figure-14}, suggesting that RSEs come from the ``computational'' subset of a subject. However, within computer science research within the UK just under 23\% of academics and researchers are female~\cite{Napier-Diversity-Report}. 
This perhaps suggests that a larger percentage of female CS undergraduates move on to a research or faculty position in CS than male undergraduates, or that female CS researchers are moving into the field after undergraduate studies in other areas. However, it could also be attributed to other factors such as industry hiring trends. Of interest here is that, while over half of RSEs have a Physics and Astronomy or Computer Science-focused first degree, the percentage of female RSEs is still somewhat below the percentage of female CS research/faculty staff.

The areas that RSEs support are more widespread than the fields they are currently recruited from. There are teams offering general RSE support at a number of institutions and many RSEs working in the biosciences, geosciences, and medicine. This suggests that a challenge research software engineering has is encouraging more people with first degrees in these areas (which have more balanced gender representation) to become RSEs. This is important because domain knowledge can be especially valuable when undertaking RSE projects. Things to consider include the wording of job adverts, which can affect who applies for them~\cite{Apa18} --- can we encourage more candidates from other disciplines? There may be lessons for the software development field more generally --- there are a significant number of RSEs whose background is in biological sciences or geosciences, as well as some from social sciences and the humanities, showing people with ``non-traditional'' degree backgrounds seek careers as software practitioners. 

One encouraging aspect of Figure~\ref{fig:rse-flow} is there appear to be many lines going from one discipline on the left hand side to a different discipline on the right hand side, representing an RSE working in an area different from the one where they trained. This suggests that 
there is scope to improve diversity within the RSE community through attracting individuals from more diverse domains into the RSE space. This provides an opportunity for more immediate improvements in the diversity of the RSE community than the multi-year timeline that we might expect for efforts to improve diversity amongst undergraduate cohorts. To investigate this movement between domains, which we call \emph{``domain mobility''}, we undertook further analysis of the most recent 2018 RSE survey data.

\begin{table}
\vspace{-4mm}
\centering
\begin{tabular}{ || m{6.6cm} | m{2.5em} | m{5em} | m{6em} || } 
 \hline
 Discipline & N & Partially mobile (\%) & Fully mobile (\%) \\
 \hline  \hline
 Biological Sciences & 95 & 44.21 & 8.42 \\
  \hline
 Chemistry & 22 & 40.91 & 22.73 \\
  \hline
 Computer Science & 241 & 51.45 & 22.41 \\
  \hline
 Electrical \& Electronic Engineering & 36 & 41.67 & 38.89 \\
  \hline
 Geography \& Environmental Sciences & 46 & 52.17 & 6.52 \\
  \hline
 Mathematics & 69 & 47.83 & 47.83 \\
  \hline
 Physics and Astronomy & 261 & 31.80 & 19.92 \\
 \hline
\end{tabular}
\caption{Example of domain mobility across a selection of disciplines from the 2018 RSE survey data where N$>$20.}
\label{table:table1}
\vspace{-8mm}
\end{table}

Table~\ref{table:table1} shows an example of “domain mobility” for a series of domains where the number of survey respondents was $>$ 20. We define \emph{partial mobility} as RSEs working in one or more domains outside the domain of their highest degree, while also still undertaking work in the domain of their degree. \emph{Full mobility} applies a further filter to partial mobility by excluding RSEs who still work in the domain of their highest degree. So, in the case of Mathematics, where the figures for partial and full mobility are the same, none of the survey respondents who said their highest degree was in Mathematics undertake RSE work in the domain of Mathematics. It should be noted that while domain mobility is a positive concept, and the ability to work across a wide variety of different domains is a great benefit, we do not have any data on the reasons for these movements between domains. The movements may be individuals wanting to undertake RSE work in, and learn about, a different domain to their original area of study. However, there are also likely to be cases of movement between domains being made out of necessity due to a lack of job opportunities in an individual's domain of choice, for example. 

\subsection{Code and Data Availability}

The Jupyter Notebooks used to perform the analysis are available from \cite{papersources}. International RSE data can be obtained from \cite{RSESurvey2018}, apart from some gender data which have not been publicly published at the time of writing.

\section{Improving EDI in the research software community}
\label{section:improving-diversity}

In this section we investigate three areas that represent both challenges and opportunities for improving equity, diversity and inclusion (EDI) within the research software community. Through the material in this section, we hope to highlight possible reasons for the lack of diversity in the research software space while also providing some thoughts and guidance on how these can be addressed and how the community can work together to help improve a wide range of aspects of diversity. The material here is aimed at both developers of research software, and at team leaders, technical managers and individuals in other management and community leadership roles. For individuals who write software, we aim to offer thoughts and advice that can help improve understanding of the benefits of diversity and recognise situations that can contribute to a lack of diversity. For managers and team leaders, we hope to improve recognition of opportunities to address diversity challenges, both locally within teams and in the wider community in the context of events and activities.

\subsection{Safety in similarity?}
\label{section:safety-similarity}

Diversity, whether in terms of gender, ethnicity, skills or other characteristics, is important in bringing different perspectives, ideas and experiences to a community. Diverse teams can lead to higher quality science \cite{Campbell2013} and improved technology business performance \cite{DiversityPerformance}. 

However, as highlighted by Merritt~\cite{Merritt18} in the context of hiring staff, individuals want to work with people who are like them and Lang \& Liu~\cite{Liang_2007} point out contradictory evidence for the benefits of diversity within teams. They highlight work by Byrne in the 1971 book ``\textit{The Attraction Paradigm}'' that suggests certain similarities within teams can be helpful in ensuring effective working environments. However, they do also highlight a series of other work that supports the idea of diversity within teams and groups being beneficial. Van der Zee, et al.~\cite{Van_Der_Zee_2004} highlight previous work that suggests that we tend to have a positive response to similarity while the opposite is true for dissimilarity. They also point out other previous work suggesting that we have an attraction to people who share similar attitudes and values to our own since this makes communication with them easier.

A very large study of over 20 years of scientific papers~\cite{Freeman_2015} shows that paper co-authors are more likely to be of similar ethnicity. However, where papers are the result of collaborations between authors at different locations, it was shown that this can result in them being published in journals with high impact factors and receiving more citations. 

The above examples demonstrate that there is a perceived ``safety'' in being around people who have similar interests, backgrounds and/or values. We see that this is likely to work both ways and that in the case of a community, such as the RSE community or groups of software practitioners more generally, individuals choose whether or not to engage partly based on whether they see people like themselves within that community. Of course, while individuals may seek out others who they feel are similar or who they have things in common with, diversity can be extremely important and hugely valuable in providing different views, ideas, attitudes and perspectives. Ultimately this can lead to important benefits, even if there may be learning experiences to be had along the way. These include, for example, getting more used to working with people who may not look at every opportunity, challenge or research problem in the same way. Many of us will have had experiences that support this when attending community events and workshops which often involve breakout group discussions or problem-solving tasks.

Awareness is an important step here and this begins with \textbf{educating team leaders, community managers and event organisers to be more aware of situations where there is a lack of diversity and the benefits of addressing this}. Developing a stronger understanding of the benefits of diversity, but also the challenges that can arise in ensuring that diverse groups can collaborate effectively, is an important step towards addressing the lack of diversity in the scientific software community. This is especially important for individuals in leadership or organisational roles. Various guides, for example the Hopper Conference Diversity Guide~\cite{HopperGuide} and NumFOCUS DISCOVER Cookbook~\cite{DISCOVERCookbook}, provide detailed information on approaches for helping to ensure diversity at conferences and events. There is also a role for RSE group leaders in encouraging and enabling the RSEs in their teams to collaborate with other institutions, as this diversity might lead to higher impact of their work.

\subsection{Increasing equity, diversity and inclusion at events}

The events that we consider in this section include everything from large conferences and workshops to small local community events with only a small number of attendees. Large events often have dedicated organising teams and potentially co-chairs specifically dedicated to areas such as EDI. This is unlikely to be the case for a small community event with perhaps only a few 10s of participants. Nonetheless, being able to widen participation at events requires you to start with an understanding of the community the event is targeted at and an awareness of the particular aspects of diversity that you aim to improve~\cite{ConferenceDiversity}. One suggestion for improving equity and inclusion at conferences, from work looking at 30 conferences in the conservation and ecology domain, is to write up and promote details of the actions and processes followed to support improving EDI~\cite{Tulloch_2020}. This work demonstrates that concerns around ensuring diversity at conferences and events are not specific to the research software community. The sort of open approach espoused through supporting and promoting EDI activities could be beneficial across many, if not all, domains. It would help to offer a demonstration of a conference's efforts to support aspects of EDI and also provide evidence for the wider community to help identify what works and what is less successful. This should help to avoid repeating less successful approaches to addressing diversity and inclusion concerns across different domains.

Ensuring that event committees, speakers and panel members reflect the diversity that organisers would like to see among an event's attendees has been effective in some fields. This builds on some of the ideas highlighted in Section~\ref{section:safety-similarity}. Research that looked at presenters across 21 meetings of the American Association of Physical Anthropologists found events that had either female, or both female and male organisers, resulted in a much higher percentage of women as first authors of presented papers and posters~\cite{Isbell_2012}.
However a study by Bano and Zowghi of six software engineering conference series~\cite{Bano_2019} found that having a female conference chair or program committee chair \textit{did not} significantly affect the number of keynote speakers that were female, or the make up of the program committee. This work did not, however, look at all conference speakers.

While the figures shown in Section~\ref{section:survey} show a lack of diversity within the RSE community, there is a perception, when attending RSE workshops and events, of gender and ethnic diversity being significantly better than the headline survey figures suggest. RSE events generally have a significant number of female speakers, workshop organisers and community leaders. We see with the Collaborations Workshop series, an event that focuses on general research software practice, a positive change in the diversity balance. It has had a policy of ensuring diversity of keynote speakers, and in 2020, 55\% of steering committee members were female, 42\% of all speakers were female, 34\% of attendees reported their gender as female, and 20\% of attendees reported their ethnicity as non-white or mixed.

Nonetheless, the challenge remains of attracting more individuals from a wider range of backgrounds to get involved with RSE. A concrete action that can be taken here is to work to \textbf{increase the diversity amongst organisers, speakers and sponsors at RSE events and be more open about the approaches taken to support this}. The latter part of this recommendation should help organisers of other events, both within the research software community and beyond, to learn from and build on efforts to improve EDI. Ultimately this should help to accelerate the process of improving diversity throughout the research community.

\subsection{An inclusive culture and safe space}

An inclusive culture is an important foundation of diversity~\cite{ChamorroPremuzicInterventions19}. 
Research teams that include RSEs frequently have only one or two individuals in RSE roles. Having valuable but different skills, and different career aims, can make RSEs feel like the ``odd one out'' within a research group or team. As a result, RSE communities of practice have developed in recent years to support individuals in this field. Possibly due to previous perceived marginalisation, they tend to be open, welcoming communities. These communities provide a great opportunity to meet, network and collaborate with others who understand the challenges RSEs face in their day-to-day work. 

At conferences and events, one way of formally promoting inclusion is having a clear and well-publicised \emph{Code of Conduct}. The importance of a Code of Conduct and the challenges it can help to address and provide guidance on are highlighted in~\cite{Favaro_2016}. While still not commonplace at traditional academic conferences~\cite{Foxx2019}, it is unusual to find an RSE conference or workshop that does not have a Code of Conduct -- learning from the experience of the open source community.  
To ensure best practice and clarity of the message provided, many events choose to build on widely accepted, open codes of conduct such as the template provided by the Geek Feminism wiki~\cite{GFWiki}, and also have a diversity statement. 

In the context of workplaces and research teams, it is also important that individuals feel that they fit in, both within the workplace environment itself and with the people based there. Cheryan et al.~\cite{Cheryan_2009} call this ``ambient belonging''. They undertook a series of studies within a computer science context looking at gender-based perspectives on the influence of different environments. This work showed that environments that had aspects that made them fit with computer science stereotypes reduced the interest of women participating in the field. Changing the environments to make them appear less like something that would be associated with computer science increased interest in the field. This provides an example of the sort of potentially small, but nonetheless significant, changes that can help to develop an inclusive culture, a welcoming environment and, ultimately, help to improve diversity. We consider that there may also be opportunities to help improve other aspects of diversity through identifying and making similarly small environmental or organisational changes in different contexts, for example to improve the experience for people with disabilities.

The wide ranging use of codes of conduct at research software events and the recognition of the importance of communities in helping to provide individuals with a place where they fit in is significant. It suggests that the research software community is developing events that are explicit in their desire to be open and diverse, with clear statements on acceptable behaviour. We see that progress is being made in developing an inclusive culture and providing environments through events and activities that potential participants feel are open and welcoming. It is of vital importance that this continues. There are two concrete recommendations from this analysis. Firstly, \textbf{within RSE groups, both team leaders and members should be aware of the importance of their working environment and how potentially small environmental changes may affect other team members' feelings of inclusion and belonging}. Secondly, in the context of community events and activities, \textbf{event organisers should look to highlight their support for diversity and inclusion through the use of a diversity statement and provide a clear code of conduct highlighting acceptable behaviour}.

The Software Sustainability Institute runs a fellowship programme recognising the diverse roles and skills of those working to promote research software practice. An evaluation of the programme showed it plays an important role in supporting communities of best practice and skills transfer, and that a significant benefit is the way it has raised the profile of software in research, and those people who develop and advocate for it~\cite{fellowsurvey2018}. This has had positive effects for those who may previously have considered themselves as `outsiders' in the role, or lacked confidence. This is exemplified by a comment from a female respondent: 

\begin{quote}``\textit{Despite getting a PhD partially from a computer science programme, I could see that my skills and knowledge were always at least to some extent dismissed or doubted [...] since being elected a SSI fellow I most definitely observed a significant drop in mansplaining... I have little doubt that the SSI fellowship was a significant [reason] I got my current position as (Head of Division at a Supercomputing Center).}''~\cite{fellowsurvey2018}
\end{quote}

\section{Discussion and Conclusions}
\label{section:conclusion}

Our research has shown that the gender diversity of RSEs is similar to the field of software engineering, but does not reflect the gender balance of the academic research workplace. This is particularly true where RSEs are working in domains such as the biosciences, geosciences and medicine. Ethnic diversity among RSEs is worse than in the wider software industry, but it is unclear why. The number of RSEs with disabilities reflects the academic research workplace, and is poor in comparison to the IT profession. Given that 19\% of the UK working age population report a disability, there is much to be done to provide a equitable workplace for RSEs with disabilities. Differences in workplace culture, environment or incentives may be factors, and further research is required. However, general interventions to improve diversity appear to be increasing gender and ethnic diversity at events.

We have highlighted the importance of communication and collaboration between individuals building software to support research and the computational scientists and researchers that they collaborate with. We have also shown that diversity among all parties in these collaborations can lead to better communication, a wider range of ideas and perspectives, and, ultimately more effective collaborations that produce higher quality outputs. The three areas discussed in Section~\ref{section:improving-diversity} as key opportunities for improving equity, diversity and inclusion within the field of research software can be summarised as follows:

\begin{itemize}
    \item \textbf{Safety in similarity?} The perceived safety of looking to collaborate and work with individuals who have similar interests, backgrounds or values is widely recognised. While there is research that suggests there can be benefits to this approach, there is also extensive work highlighting the benefits of diversity. However, this is something that can be approached from two different perspectives and individuals may be more likely to join and engage with a community if they see people like themselves within that community.
    \item \textbf{Increasing diversity at events:} It is important to highlight and promote actions being take to improve EDI in the context of events. It is also important to ensure that both organising groups and speakers reflect the level of diversity that an event's organisers would like to see among its participants.
    \item \textbf{An inclusive culture and safe space:} This is a very important aspect of both participating in a community, and being part of an RSE group, for everyone involved. A well-publicised Code of Conduct and diversity statement with clearly defined processes supporting them are key elements in helping to ensure this within the context of events. In RSE groups, leaders and team members need to be aware of the importance of an inclusive culture and working environment.
\end{itemize}

Poor diversity in the community of research software developers is likely a result of the low levels of diversity in the disciplines that currently form the major path towards an RSE role or career. We have identified scope to widen the range of areas from which RSEs are recruited and it is hoped that this can be achieved by better use of language in the way that RSE roles are advertised and in the way that RSE is promoted, more generally. This may also have relevance for professional software practitioners looking to get involved in the research community, as it is clear that there are many people with ``non-traditional'' degrees seeking careers developing software. There is a role for professional bodies, such as the British Computer Society, Association for Computing Machinery and IEEE Computer Society, to support and embrace these career paths and help improve the gender diversity at undergraduate and postgraduate level.

Nonetheless, we feel that despite the knowledge gained from our extended analysis of existing survey data, and the pre-existing material that we have referenced, this is a hugely complex but very important area. It would benefit extensively from additional evidence that could be gathered through a range of further empirical studies. Some of the areas that may be considered most important for such studies in the short term include:

\begin{itemize}
\item Exploring why levels of diversity among RSEs are lower than in many of the areas of study and research that already feed into RSE careers, such as Computer Science and Physics and Astronomy. What are the levels of diversity in other career paths for these subjects? What influences the career choices that individuals make and are there specific aspects that steer them away from an RSE career?
\item Trialing approaches for increasing diversity and inclusion at events, including workshops and conferences. Gathering statistics on the relative success of these approaches and their contribution to improving diversity within the computational science developer community.
\item Looking at opportunities to further increase ``domain mobility'' through taking advantage of existing courses/training material or offering new information that can help individuals move more practically between working in different domains. Look at take-up of such opportunities across different domains and use this as a basis for longer-term analysis on how diversity in the RSE community changes over time.
\end{itemize}

Once individuals become RSEs and engage with the RSE community, the RSE community meets many of the requirements to ensure that it welcomes and supports diversity. The open, inclusive, nature of the community is of great importance here and, using gender diversity as an example, there are already several women in highly-visible leadership roles.

Developing software for science and the wider computational research domain is challenging. The relatively new field of research software engineering encompasses many of the individuals undertaking these software development tasks, regardless of whether their official job title considers them to be a researcher, professor, software engineer or RSE. It is clear that research software engineering, and research software in general, shares many diversity challenges with the wider software development field but we feel there are many opportunities to address this. This begins with increasing awareness of the different aspects raised in this paper and using this as a basis to develop the environment and opportunities to help build equity, diversity and inclusion within the community. We hope that the research software community continues to grow, and become more diverse by sharing and learning from other practitioners.  

\section{Acknowledgement}

We thank Simon Hettrick, James Graham and Rob Haines for their input, and Olivier Philippe who originally processed the international RSE survey data. NCH and CJ acknowledge support from EPSRC / BBSRC / ESRC / NERC / AHRC / STFC / MRC grant EP/S021779/1 for the UK Software Sustainability Institute. JC acknowledges support from EPSRC grant EP/R025460/1.

\bibliographystyle{splncs04}
\bibliography{RSE-diversity} 


\end{document}